\documentclass[aps,prl,twocolumn,showpacs,amsmath,a4paper,amssymb,10pt]{revtex4-1}
\usepackage{graphicx}
\usepackage{CJK}
\usepackage{epstopdf}

\usepackage{amsmath,amssymb,MnSymbol}
\usepackage{color}
\usepackage{lipsum}

\newcommand{\wpcm}{W/cm$^2\,$}
\newcommand{\arp}{Ar$^+\,$}
\newcommand{\arpp}{Ar$^{2+}\,$}
\newcommand{\arps}{Ar$^{+*}\,$}
\newcommand{\ars}{Ar$^{*}\,$}

\begin{document}
\begin{CJK*}{UTF8}{gbsn}

\title{Frustrated Double Ionization of Argon Atoms in Strong Laser Fields}

\author{Seyedreza Larimian}
\author{Sonia Erattupuzha}
\affiliation{Photonics Institute, Technische Universit\"at Wien, A-1040 Vienna, Austria, EU}
\author{Andrius Baltu\v{s}ka}
\author{Markus Kitzler}
\author{Xinhua Xie (谢新华)}
\email[Electronic address: ]{xinhua.xie@tuwien.ac.at}
\affiliation{Photonics Institute, Technische Universit\"at Wien, A-1040 Vienna, Austria, EU}
\pacs{33.80.Rv, 42.50.Hz, 82.50.Nd}
\date{\today}

\begin{abstract}
We demonstrate kinematically complete measurements on frustrated double ionization of argon atoms in strong laser fields with a reaction microscope. We found that the electron trapping probability after strong field double ionization is much higher than that after strong field single ionization, especially in case of high laser intensity. The retrieved electron momentum distributions of frustrated double ionization show a clear transition from the nonsequential to the sequential regime, similar to those of strong field double ionization. The dependence of electron momentum width on the laser intensity further indicates that the second released electron has a dominant contribution to frustrated double ionization in the sequential regime.
\end{abstract}

\maketitle
\end{CJK*}


When an atom or a molecule is exposed to a strong laser field it may become singly or multiply ionized~\cite{Fittinghoff1992,Krausz2009}. After the strong field interaction, a fraction of the ionized electron wave packets with near-zero kinetic energies can be trapped into high-lying Rydberg states, a process also known as frustrated field ionization \cite{Nubbemeyer08,Manschwetus2009,Ulrich2010,Wu2011,liu12,Wolter14,li14,Larimian2016,Larimian2017}. Frustrated double ionization (electron trapping after double ionization) has been experimentally studied for small molecules, including H$_2$ and argon dimers, using Coulomb explosion imaging \cite{Manschwetus2009,Ulrich2010,Manschwetus2010,Wu2011,Emmanouilidou2012,Cheng2017} and theoretically using the classical trajectory Monte Carlo method \cite{Shomsky2009,Chen2016}.
In these experiments the trapping process is identified by the kinetic energy released (KER) during the Coulomb explosion of the molecules. Since an electron trapped in high-lying Rydberg states does not fully shield the nuclear charge, the KER for a molecule with an electron in Rydberg states is higher than that for a non-excited molecule. Since this method is based on the measurement of KER from molecules undergoing Coulomb explosion, it is applicable to neither atomic targets nor molecules that do not fragment.

In this paper, using an alternative method developed in our previous work~\cite{Larimian2016}, we report on kinematically complete experiments of electron trapping processes during strong field double ionization of argon atoms. Strong field double ionization may happen sequentially, where the two electrons are removed one after another by the laser field, or non-sequentially, where the second electron is released during the recollision of the first electron with the parent ion~\cite{Larochelle1998,Weber2000,Becker2012}.
We show that the trapping probability is strongly enhanced in the sequential ionization regime. Based on our experimental data we explain the electron dynamics underlying these observations.

\begin{figure*}[htbp]
 \begin{center}
  \includegraphics[width=\textwidth]{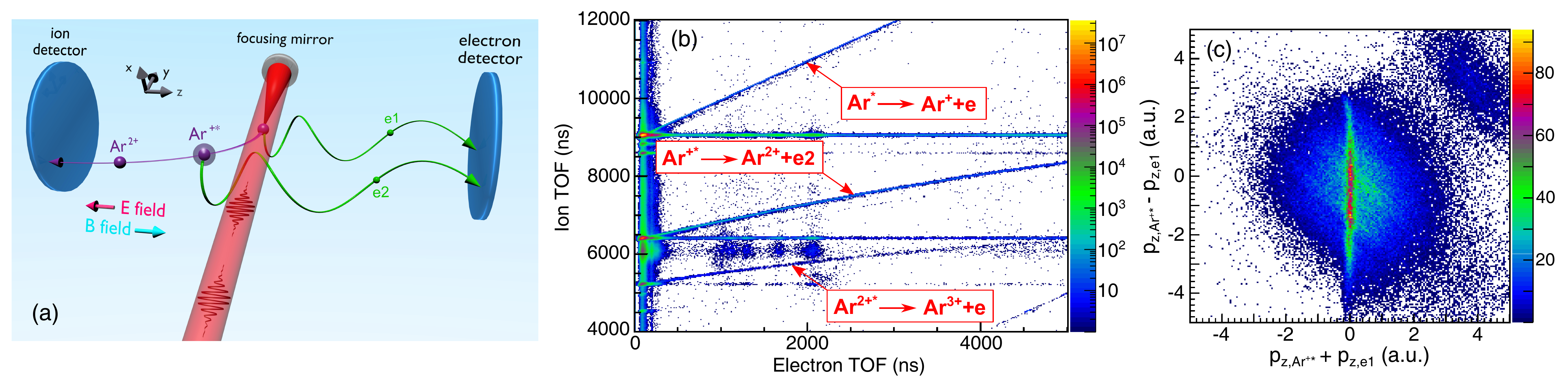}
  \caption{(a) Schematic view of the experimental setup and the three-body detection of the dc field or BBR ionization of Ar$^{+*}$ from strong laser field interaction in the laser focus. (b) Measured photo-electron-photo-ion coincidence distribution shows the relation between the detected electrons and ions. The lines represent the correlated signal of electrons and ions from the dc field or BBR ionization of Ar$^{*}$, \arps and Ar$^{2+*}$ are labeled. (c) Correlation between the retrieved momentum of Ar$^{+*}$ and electrons released by strong laser fields along the laser polarization direction. Laser peak intensity is $5\times10^{14}$ \wpcm.} \label{fig1}
 \end{center}
\end{figure*}

In our experiments we employed a reaction microscope \cite{doerner00,ullrich03} for three-body coincidence detection of two electrons and their parent ion created during the interaction of argon atoms with strong laser pulses (Fig.~\ref{fig1}(a)). Laser pulses linearly polarized along the spectrometer axis ($z$-direction) were provided by a home-built Titanium:sapphire laser amplifier system. The pulses had a center wavelength of 790 nm, a pulse duration of 25 fs and peak intensities in the range of 10$^{14}$ to 10$^{15}$ W/cm$^2$. A weak homogeneous dc field of a few V/cm is applied along the $z$-direction. This field not only accelerates charged particles to the detectors but also induces field ionization of high-lying Rydberg states populated during the strong field interaction~\cite{Larimian2016}. Additionally, a homogeneous magnetic field of 12.3 gauss ensures 4$\pi$ detection of electrons from strong field interaction. More details on the experimental setup can be found in our previous publications \cite{xie12,xie12prl,Larimian2016}.

Previously, we developed a method to characterize Rydberg states in atoms and molecules formed during strong field interaction. This method employs coincidence detection of Rydberg electrons released by tunnel ionization in the spectrometer dc field and single photon ionization by blackbody radiation (BBR)~\cite{Larimian2016,Larimian2017}. During the strong field interaction, one electron ($e1$) is removed through strong field ionization and a second electron (($e2$)) is trapped in high-lying Rydberg states to form Ar$^{+*}$. The detection of frustrated double ionization of an argon atom is depicted in Fig.~\ref{fig1}(a).
During the flight of the excited ion (Ar$^{+*}$) to the detector, the spectrometer field or BBR further releases the Rydberg electron ($e2$). We recorded the time-of-flight (TOF) and position information of all three particles (\arpp and two electrons) with two multi-hit delay-line anode detectors. As shown in Fig.~\ref{fig1}(b), the correlated signals of Rydberg electrons with their parent ions (Ar$^{2+}$) represent a long parabolic curve in the photo-electron-photo-ion coincidence (PEPICO) distribution.
The PEPICO distribution also contains correlated signals of Rydberg electrons with \arp and Ar$^{3+}$ originating from electron trapping after strong field single and triple ionization, as indicated in Fig.~\ref{fig1}(b). We note that Rydberg electrons correlated with dications were also observed in separate measurements with neon atoms and acetylene molecules.

With the measured TOF and position data on the detection of Rydberg electrons and Ar$^{2+}$, we retrieved the emission time ($T$) of the Rydberg electron and the momentum vector of the Ar$^{+*}$ produced during the strong field interaction. With the assumption of gaining negligible energy during the removing of the Rydberg electron, we calculate the emission time $T$ of the Rydberg electron from the acceleration function of the \arps in the dc spectrometer field which yields a relation of:
\begin{equation}\label{eq1}
t_{e2}=T+\sqrt{T^2{m_e}/{m_{Ar}}+t_{e0}^2}
\end{equation}
where $t_{e2}$ is the TOF of the Rydberg electron, and $m_e$ and $m_{Ar}$ are the electron mass and the mass of an argon atom, respectively, $t_{e0}=\sqrt{\frac{2m_eL_e}{E_{dc}q}}$ is the TOF of a zero-momentum electron released during strong field ionization determined by the spectrometer electric field $E_{dc}$, $L_e$ is the distance from the laser focus to the electron detector, and $q$ denotes the electron charge. From Eq.~\ref{eq1} we derive the emission time $T$ as a function of the measured $t_{e2}$,
\begin{equation}\label{eq:t2}
T= t_{e2}-\sqrt{t_{e0}^2+(t_{e2}^2-t_{e0}^2)m_e/m_{Ar}}.
\end{equation}
With the retrieved survival time $T$, we can calculate the momentum of the \arps gained in the laser field along the laser polarization direction, which yields:
\begin{equation}
p_{z,Ar^{+*}}=E_{dc}(0.5t_{r0}^2-0.5T^2+t_rT-t_r^2)/t_r,
\end{equation}
where $t_{r0}$ and $t_r$ are TOFs of \arp with zero initial momentum and \arpp from ionization of \arps, respectively.


With the retrieved momentum of \arps [from the measured data of \arpp and the Rydberg electron ($e2$)], we can check the quality of the three-body coincidence detection of \arpp with two electrons for frustrated double ionization process. In Fig.~\ref{fig1}(c) we show the momentum correlation between the \arps and the electron ($e1$) released during the strong field interaction with the sum and difference of their momenta. Due to momentum conservation, the momentum sum of the strong-field electron ($e1$) and \arps from the same atom is close to zero with a narrow momentum distribution determined by the initial momentum of the argon atom. The correlation in Fig.~\ref{fig1}(c) shows that we achieved a high confidence in the three-body coincidence detection of two electrons and their parent ion \arpp. For further data analysis, we applied the coincidence selection condition of $\left|p_{z,Ar^{+*}}+p_{z,e1}\right|<0.1$ a.u. for frustrated double ionization to minimize the false coincidence rate.


\begin{figure}[htbp]
 \begin{center}
  \includegraphics[width=0.45\textwidth]{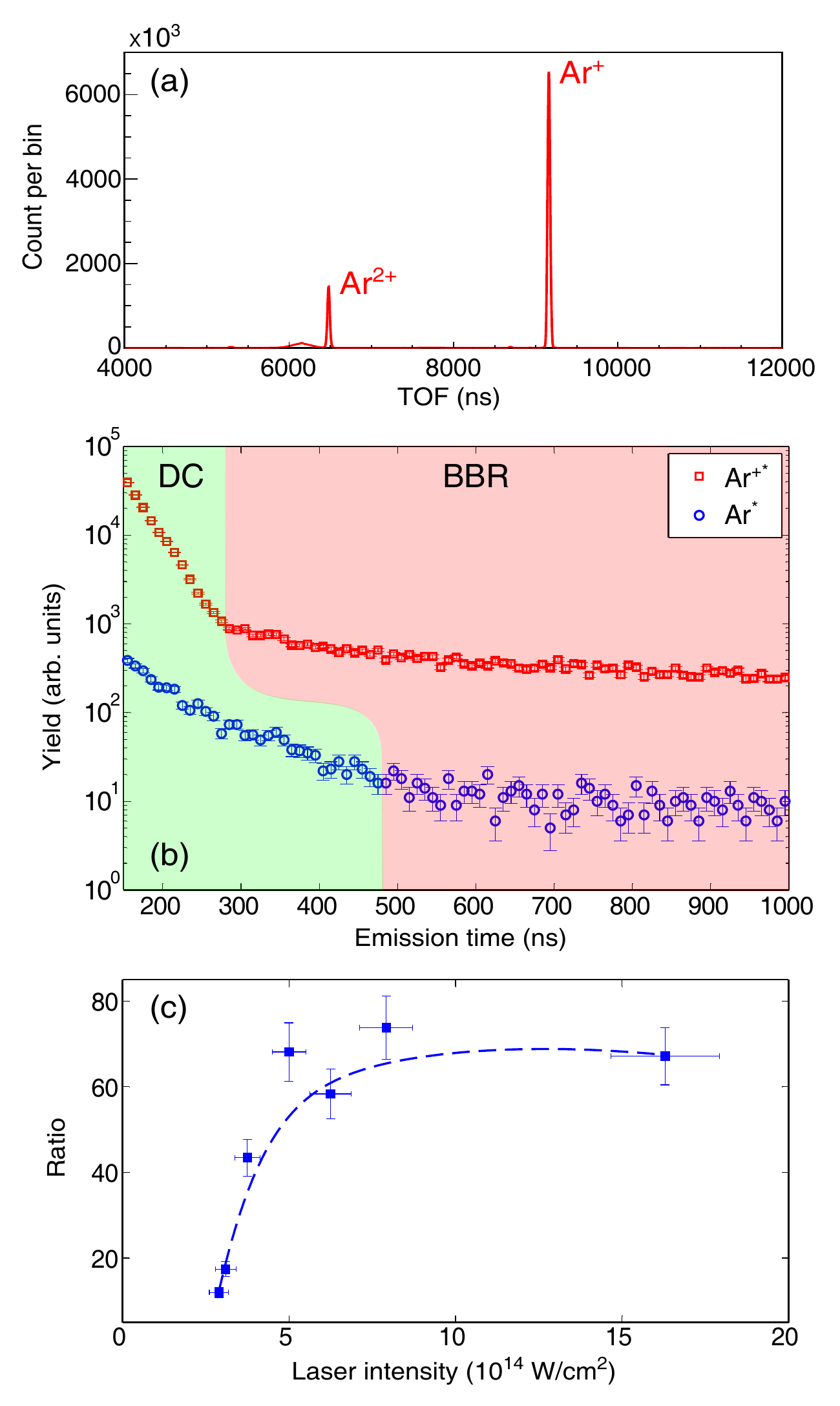}
  \caption{(a) Measured time-of-flight spectrum of photoions. (b) Distributions of measured emission time of Rydberg electrons from \ars (blue circles) and \arps (red squares) for laser peak intensity of $7.8\times10^{14}$ \wpcm and the dc field strength of 3 V/cm. (c) The ratios of trapping probability between the single and double ionization processes as a function of laser peak intensity. The blue dashed line is a fitting curve to guide the eyes.} \label{fig2}
 \end{center}
\end{figure}

Figure~\ref{fig2}(b) presents measured signals of Rydberg electrons from \ars and \arps over emission time for a laser peak intensity of $7.8\times10^{14}$ \wpcm and a dc field strength of 3 V/cm. As reported in our recent publication \cite{Larimian2016}, the yield of Rydberg electrons contains two contributions: due to ionization in the dc spectrometer field and due to BBR-induced photoionization. Dc field ionization contributes dominantly to the signal with small emission times and causes a fast decay with an ionization rate of about $10^{-2}$ ns$^{-1}$, while BBR-induced photoionization is responsible for the slowly decaying signal with an ionization rate of about $5\times 10^{-4}$ ns$^{-1}$.
In addition to the different decay rates, another clear observation is that the Rydberg electron signal from \arps is much stronger than that from \ars, even though in measurement the strong field single ionization probability (\arp) is about 4 times higher than the strong field double ionization probability (\arpp), which is shown in Fig.~\ref{fig2}(a). To compare the electron trapping probability in the single and double ionization processes, we calculated the BBR-induced ionization signal ratio between the Rydberg electrons from \ars and \arps, taking into account the ratio between the signals of strong field single and double ionization. This ratio as a function of laser peak intensity is shown in Fig.~\ref{fig2}(c).
The electron trapping probability after double ionization is more than one order of magnitude higher than that after single ionization. The trapping probability ratio increases with the laser intensity in the range from 2$\times 10^{14}$ \wpcm to 1$\times 10^{15}$ \wpcm and saturates at about 8$\times 10^{14}$\wpcm. One reason for the higher trapping probability in double ionization is obviously that two electrons can contribute to the trapping probability. The other reason for the higher trapping probability for double ionization might be the different shape of the Coulomb potentials.
The Coulomb potentials are $-1/r$ and $-2/r$ for the electron trapping process during single and double ionization, respectively. The Coulomb potential influences the trapping probability in two ways. First, the potential $-2/r$ is deeper than $-1/r$, which leads to a larger trapping volume, allowing more low kinetic electrons to be trapped. The spatial volume of the potential $-2/r$ for trapping electrons with a certain near-zero kinetic energy is 8 times as that of the potential $-1/r$. The second effect is the Coulomb focusing effect \cite{Brabec1996,Comtois2005}, which is also stronger for the potential $-2/r$ than $-1/r$. The Coulomb volume and focusing effects together with the doubled trapping probability lead to the strongly increased electron trapping probability after double ionization than after single ionization. The dependence of the ratio on the laser intensity will be further discussed later in the text.

\begin{figure}[htbp]
 \begin{center}
  \includegraphics[width=0.45\textwidth]{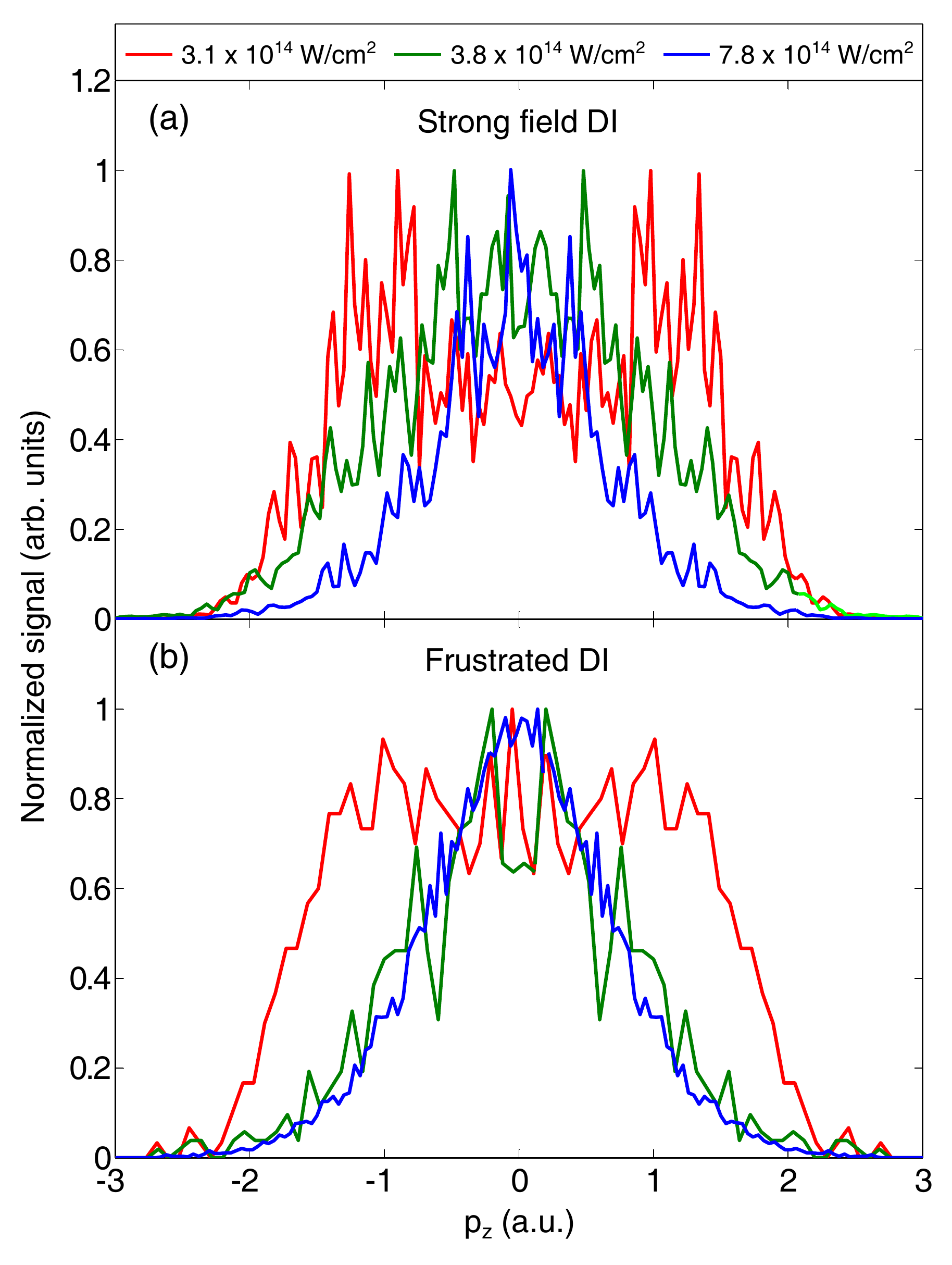}
  \caption{Measured electron momentum distributions along the laser polarization direction from strong field double ionization (a) and frustrated double ionization (b) for laser peak intensities of 3.1, 3.8 and 7.8$\times10^{14}$ \wpcm.} \label{fig3}
 \end{center}
\end{figure}

Since the electron's final momentum is determined by the vector potential of the laser field at the electron releasing time, measured electron momentum distributions contain temporal information on the strong field interaction. We present electron momentum distributions along the laser polarization direction from strong field double ionization and frustrated double ionization for three different laser peak intensities (3.1, 3.8 and 7.8$\times10^{14}$ \wpcm) in Fig.~\ref{fig3}(a) and (b), respectively. The sharp peaks in the momentum distributions are ATI-like structures caused by the interference of electron wave packets released at different times during the laser pulse \cite{Parker2006,arbo10,xie12interference}.
Figure~\ref{fig3}(a) shows that the electron momentum distribution of strong field double ionization gradually changes from a double-hump structure at 3.1$\times10^{14}$ \wpcm to a single-hump distribution at 7.8$\times10^{14}$ \wpcm as the laser intensity increases. This behaviour is a manifestation of the change of the double ionization mechanism from the non-sequential to the sequential regime \cite{Weber2000,Lein2000,Zrost2006}. When the laser intensity is low, the sequential double ionization rate is low and the recollision induced double ionization dominates. Sequential double ionization becomes the dominant mechanism when the laser intensity becomes strong enough \cite{Becker2012}. The momentum distributions of the released electron correlated with frustrated double ionization, shown in Fig.~\ref{fig3}(b), feature a similar behavior as those from strong field double ionization in Fig.~\ref{fig3}(a).

\begin{figure}[t!]
 \begin{center}
  \includegraphics[width=0.45\textwidth]{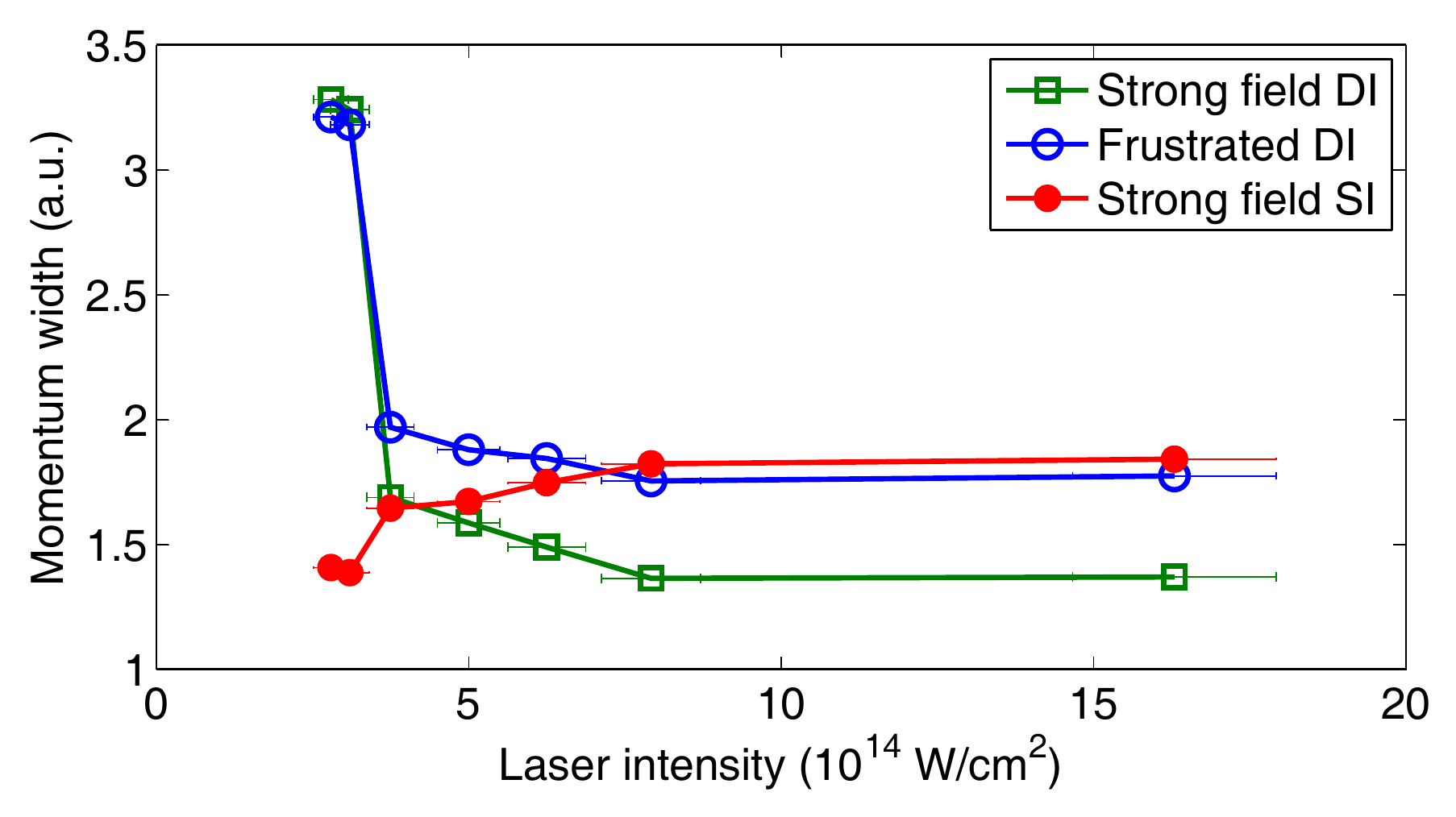}
  \caption{The width of electron momentum distribution as a function of laser peak intensity for strong field single ionization (filled red circles), strong field double ionization (green squares) and frustrated double ionization (open blue circles).} \label{fig4}
 \end{center}
\end{figure}

For a quantitative comparison of the measured momentum distributions in Fig.~\ref{fig3}(a) and (b) we retrieved their widths at the half maximum. They are plotted as a function of laser peak intensity in Fig.~\ref{fig4} together with the widths of the electron momentum distributions caused by strong field single ionization. With increasing laser intensity the widths of momentum distributions attributed to strong field double ionization decrease and level off at a peak intensity of about $8\times10^{14}$\wpcm (green squares in Fig.~\ref{fig4}). Such behavior has been reported previously and is due to the transition from nonsequential double ionization to sequential double ionization as the laser intensity increases \cite{Rudenko2008}.
On the other hand, the widths of electron momentum distributions attributed to strong field single ionization show opposite dependence on the laser intensity. They increase with increasing laser intensity and level off at an intensity of about $8\times10^{14}$\wpcm (filled red circles in Fig.~\ref{fig4}). This signifies the saturation of single ionization at a laser intensity around $8\times10^{14}$\wpcm. Before saturation, the momentum width is proportional to the peak vector potential ($\sqrt{I_0}/\omega_0$) of the laser field \cite{milovsevic2006}, which increases monotonically with the laser intensity $I_0$.

Turning to the widths of the momentum distributions of electrons released during the frustrated double ionization process (open blue circles in Fig.~\ref{fig4}), we observed a similar dependence on the laser intensity as for those electrons released by strong field double ionization. Two important facts can be inferred from a comparison of the momentum widths of electrons connected to frustrated double ionization and strong-field single and double ionization, respectively: i) At a laser peak intensity of about $3\times10^{14}$\wpcm the momentum width gained during frustrated double ionization is almost the same as that gained during strong field double ionization; ii) For a laser peak intensity higher than $8\times10^{14}$\wpcm electron momentum widths gained during frustrated double ionization and strong field single ionization are very similar.

For a laser peak intensity of about $3\times10^{14}$\wpcm, strong field double ionization happens dominantly nonsequentially: One electron is released at the peak of the laser field and subsequently accelerated in the laser field. When the laser field changes sign, the released electron can be driven back to the parent ion and collide with it, thereby kicking out a second electron \cite{Fittinghoff1992}. In nonsequential double ionization the momentum of the two indistinguishable electrons shows strong correlation which manifests a double peak structure in the momentum distribution \cite{Rudenko2007,Staudte2007}. In frustrated double ionization, one of the two electrons is trapped after the conclusion of the laser pulse. The non-zero momenta of the two electrons lead to the suppression of the electron trapping probability because only electrons with near-zero kinetic energy can be trapped by their parent ions.

For laser peak intensities higher than $8\times10^{14}$\wpcm, strong field double ionization is dominated by a sequential emission dynamics. In the sequential double ionization process, the first electron is released at the leading edge of the laser pulse and the second ionization occurs preferentially around the peak of the laser pulse. For such high intensities single ionization becomes saturated. Since the second ionization potential (27.63 eV for Ar) is much higher than that of the single ionization (15.76 eV for Ar), according to the ADK theory \cite{ADK} the second ionization step happens in a much narrower time window within a half cycle of the laser field. As a consequence, the momentum width of the second electron is smaller than that of the first electron.
As in the measurement we cannot distinguish between the first and second electrons due to the overlapping of their TOF-distribution, the measured electron momentum distribution contains contributions from both electrons. Thus, the measured electron momentum width of strong field double ionization would be smaller than that of the first electron and bigger than that of the second electron, which explains the smaller momentum width of strong field double ionization than that of strong field single ionization in the high intensity region.

Finally, we return to the momentum width of electrons released by frustrated double ionization. Fig.~\ref{fig4} shows that they are close to those of electrons released by strong field single ionization in the high intensity region. This experimental observation is strong evidence that in the sequential double ionization region mainly the second electron becomes trapped and that the measured momentum width is mainly due to the first ionization step. This can be explained by the quantum diffusion of electron wave packets before trapping. Since the first electron is released earlier and spreads for longer time in the continuum than the second electron until the conclusion of the laser pulse, the first electron wave packet diffuses more severely in space. This leads to a strong suppression of the trapping probability in comparison with the second electron. In addition, because of the smaller lateral momentum of the electron released during the second ionization step from the larger ionization potential \cite{Arissian2010}, the diffusion velocity of the second electron in space is smaller than that of the first electron. This also contributes to the higher trapping probability of the second electron than that of the first electron.

The transition from the nonsequential to the sequential double ionization scenario also explains the dependence of the trapping probability ratio on the laser intensity, shown in Fig.~\ref{fig2}(c). In the low intensity region, the trapping probability in double ionization is less enhanced due to electrons with pronounced momentum offset induced by recollision. With increasing laser intensity more and more double ionization events take place sequentially. This leads to the experimentally observed increase of the electron trapping probability after double ionization.


In conclusion, we studied experimentally electron trapping processes during strong field double ionization of argon atoms using three-body coincidence detections. We observed a strongly enhanced trapping probability after strong field double ionization of argon atoms in comparison with single ionization. The measured intensity dependence of this enhancement indicates that in the sequential double ionization regime the trapping process is dominated by the second detached electron because of its smaller quantum diffusion as compared to the first electron. In the nonsequential double ionization regime we find that the trapping probability is strongly suppressed as compared to that in the sequential double ionization regime. We attribute this to the strong correlation between the two electrons which results in momentum distributions offset from zero.

The here applied method of coincidence detection of Rydberg states can be also used for studying electron trapping processes in triple or higher-order ionization processes, and for the electron impact excitation process as well~\cite{moiseiwitsch1968electron}. Moreover, this method can be directly applied to experiments on electron trapping in laser-induced molecular ionization and dissociation processes using multi-particle coincidence detection. This may trigger new research on the impact of Rydberg electrons on laser-induced molecular reactions and the interaction between Rydberg atoms and/or molecules \cite{Zhou2012,Bogomolov2014,jochim2017}.

We thank Chii-Dong Lin, Christoph Lemell and Joachim Burgd\"orfer for fruitful discussions. This work is financed by the Austrian Science Fund (FWF) under P25615-N27, P30465-N27, P28475-N27, special research programs SFB-049 Next Lite.

\end{document}